\documentstyle[12pt,aaspp4,epsfig]{article}

\def\lsim{{{}_{{}_<}^{~}\atop {}^{{}^\sim}_{~}}}
\newcommand{\be}{\begin{equation}}
\newcommand{\ee}{\end{equation}}
\newcommand{\lab}[1]{\label{#1}}
\newcommand{\rr}[1]{~(\ref{#1})}

\begin{document}

\title{Correcting Parameters of Events Based on the Entropy of Microlensing 
Ensemble}

\author{Piotr Popowski\altaffilmark{1,2} and Charles Alcock\altaffilmark{1,3}}
\altaffiltext{1}{Lawrence Livermore National Laboratory, Livermore, CA
94550, USA.}
\altaffiltext{2}{Max-Planck-Institut f\"{u}r Astrophysik,
Karl-Schwarzschild-Str.\ 1, Postfach 1317, 85741 Garching b.\
M\"{u}nchen, Germany.\\
 E-mail: {\tt popowski@mpa-garching.mpg.de}}
\altaffiltext{3}{Department of Physics and Astronomy, University of
        Pennsylvania, Philadelphia, PA 19104-6396, USA.\\
        E-mail: {\tt alcock@hep.upenn.edu}}

\begin{abstract}

We entertain the idea that robust theoretical expectations can become
a tool in removing hidden observational or data-reduction
biases. We illustrate this
approach for a specific problem associated with gravitational
microlensing.
Using the fact that a group is more than just a collection of
individuals, we derive formulae
for correcting the distribution of the dimensionless impact parameters of 
events, $u_{\rm min}$. We refer to the case when undetected biases in the
$u_{\rm min}$-distribution can be alleviated by multiplication of
impact parameters of all events by a common 
constant factor. We show that in this case the general maximum likelihood 
problem of solving an infinite number of equations reduces to two constraints, 
and we find an analytic solution. 
Under the above assumptions, this solution represents a state in which the 
``entropy'' of a microlensing ensemble is at its maximum, that is, the 
distribution of $u_{\rm min}$ resembles a specific, theoretically
expected, box-like distribution 
to the highest possible extent. We also show that this technique does not 
allow one to correct the parameters of individual events on the event by 
event basis independently from each other.

Subject Headings: dark matter --- gravitation --- gravitational lensing --- methods: statistical --- stars: fundamental parameters (luminosities, masses)
\end{abstract}

\section{Introduction}

There are two complementary processes that have led to progress in 
physical sciences: 1) new hypotheses have triggered experiments
which either verify or falsify these hypotheses, 2) observations of
unexpected phenomena force theorists to refine old or invent new models
and mathematical descriptions. Most of the achievements in astrophysics
have followed the second pattern. Here we argue that an extreme
variation of the first path can be very useful in some astrophysical problems.
We present the idea that robust theoretical expectations can become a tool
in removing hidden observational biases. We illustrate this approach by
deriving possible corrections to the distribution of impact parameters
for a sample of microlensing events.

Over the last decade, the microlensing surveys have grown from a fascinating
idea entertained by a group of enthusiastic theoreticians (Paczy\'{n}ski 1986;
Paczy\'{n}ski 1991; Griest 1991) to a reality of data sets containing tens 
and hundreds of real events (Alcock et al.\ 2000; Popowski et al.\ 2000; 
Udalski et al.\ 2000).
The more data we gather, the more we appreciate the ability to do precise
microlensing analyses. The effects of 'parallax' 
(Gould 1992; Alcock et al.\ 1995), 
binary caustic crossing (Paczy\'{n}ski \& Mao 1991; Afonso et al. 2000), 
and finite source (Gould 1994; Alcock et al.\ 1997) 
are particularly appreciated due to their ability to break degeneracies 
present in the simplest cases and provide useful constraints on stellar 
physics. 
On the other hand, blend fits (needed to determine how much flux
of a few unresolved stars at a given location has been microlensed)
are a standard tool in investigating crowding/seeing biases common to most 
of the events.
As long as one tries to infer the geometry
from the fluxes themselves,
the determinations are sensitive to the sky level and other weakly controlled
factors.
Therefore, it is possible that even doing the best possible analysis on
reported events, one still has to correct an entire data set for 
undetected biases. 

Here we argue that it is possible to obtain a more
accurate determination of the parameters of events using the information
that they belong to a microlensing family.
The robust prediction of microlensing is that different impact 
parameters are equally likely, and, as a result, the distribution
of the impact parameter is box-like with a values bracketed by 0
and $u^{*}_{\rm min}$, where $u^{*}_{\rm min}$ is the maximum value of 
$u_{\rm min}$ allowed by the minimum amplification chosen for a particular 
event selection (see Figure 1 and the following sections). 
This robust prediction may appear to break for blended events if blending
tends to populate certain $u_{\rm min}$ ranges on the expense of the others.
Similarly, it may seem to break if observing strategy, conditions or 
instrumentation preferentially select certain $u_{\rm min}$ ranges.
In the following treatment we will assume that such effects are either
insignificant or that they have been reduced to a
negligible level through some correction procedures (e.g., deblending).
Therefore, we concentrate on hypothetical irremovable biases that may originate
at the stage of data reduction and analysis.
We suggest that, when dealing with a clean (almost clean) microlensing
sample\footnote{The assumption of clean microlensing sample is
discussed in detail in \S 5.}, one should correct all individual $u_{{\rm min}, i}$ in such a way 
as to achieve the highest possible agreement between the observed and 
theoretically predicted cumulative distribution of $u_{\rm min}$.
The corrected $u_{{\rm min}, i}$ values may be used to re-determine
the duration of microlensing events. This, in turn, would lead to a new
estimate of the microlensing optical depth and modification of most likely
lens' masses.

In the most general case, one would like to use a Kolmogorov test
to choose a set of  $u_{{\rm min}, i}$, which is consistent with microlensing
at the highest possible confidence level.
The distribution function constraint coming from the Kolmogorov test is
formally equivalent to an infinite number of constraints on all the moments
of the distribution. This does not mean that the information contained 
in Kolmogorov statistic (see \S 4) is equivalent to the information contained
in all the moments of the distribution. However, the determination
of the Kolmogorov statistic may require the knowledge of all the moments
of the distribution.  
The moments are easier to deal with in the general 
approach of maximum likelihood which we are going to invoke here.
The high moments of the distribution
(even the shape parameters of skewness and kurtosis) are very weakly
constrained in the case of small number statistics.
One may think that it should be advantageous to use just lower, 
well-constrained moments. This is not always correct. Despite the fact that
higher moments are correlated with lower moments, the infinite number of 
very weak constraints coming from higher moments may overcome the statistical
signal of the well-constrained mean and variance of the distribution. 
Therefore, in the general case the limited moment approach to correct 
microlensing parameters may be biased in an unpredictable way.
However, there is one type of correction which is completely determined
just by the mean, $\mu$, and variance, $\sigma^{2}$, of the 
$u_{\rm min}$-distribution.
In this case described in \S 3, it is mathematically more elegant and
convenient to use the moment formalism as an ersatz for the complete Kolmogorov
statistic.

The structure of this paper is the following. In \S 2, we briefly review the
basic principles of microlensing and the possible biases in the determination
of stellar parameters. In \S 3, we present our new formalism
pointing to its likely practical application.
In \S 4 we discuss the question of how to obtain microlensing constraints from 
the traditional Kolmogorov procedure in the general case.
We argue that it is not possible to improve the parameter determination
of the fits to the individual highly-degenerate blending events.
Finally, in \S 5, we summarize our results.

\section{General microlensing}

The microlensing of stellar light is produced when a massive object (e.g.,
a star) passes very close to the line of sight between the source of light
and the observer (see e.g., Gould 2000 for a theoretical review). The light is 
gravitationally deflected and the processed
image changes its surface as projected on the plane of the source.
Because during lensing the surface brightness is conserved, the observed flux 
from the source
increases proportionally to the change in the image surface. As a result
the source is magnified and the maximum magnification $A_{\rm max}$ depends on 
the impact parameter $u_{\rm min}$, which describes the level of alignment 
between the source, lens and observer. The centroid of the new distribution 
of light in the sky changes, but because the shift is $\lsim 100 
\, \mu{\rm arcsec}$ for a typical event (Boden, Shao, \& van Buren 1998), 
it cannot be observed from the ground with current instruments. Therefore, 
the classical signature of the simplest microlensing is a characteristic 
achromatic light curve, which is time-symmetric with respect to the epoch 
with the highest flux.

Microlensing surveys typically have noisy photometry and therefore
are forced to allow only the events with 
$A_{\rm max}$ exceeding 1.5 or so. The minimum recorded maximum amplification,
$A_{\rm max}^{*}$, sets the limits on the largest recorded impact parameter 
$u^{*}_{\rm min}$:
\be
u^{*}_{\rm min} = \sqrt{-2 + \frac{2 \, A_{\rm max}^{*}}{\sqrt{{A_{\rm max}^{* \; 2}} - 1}}}. \lab{umin}
\ee
The stellar systems in which microlensing is observed contain enough
objects on the random enough orbits so that all the values of $u_{\rm min}$
in the range between $0$ and $u^{*}_{\rm min}$ are equally likely.

There are several effects that complicate this simple picture.
First of all, there are binary lenses, which produce lines of formally
infinite magnification (caustics) and light curves with a zoo of
shapes (Griest \& Hu 1992). Here we mean both stellar binaries
as well as stars with planetary companions. Second, the sources are not
point like, and as a result the magnification pattern differs from the 
basic case. Third, the constraints coming from the survey instruments,
sampling and site weather introduce a whole class of biases. 
Most of them can be summarized in two categories: blending and efficiency.
Blending is mostly about atmospheric seeing.
All microlensing surveys have to deal with seeing
of about 1 arcsecond or more. The seeing disk of that size typically covers 
more than 1 star in the sky, especially when one wants
to account for the faint end of the luminosity function. However, in almost
all cases there is only one star (or stellar system like a binary) that 
is microlensed. One has to make ``blend fits'' which determine
what fraction of the light observed in the seeing disk has been
actually microlensed.
Efficiency is about the duration of the survey, the frequency of sampling the
light curves and the photometric response of the system. The events shorter
than an average interval between observations are likely to go unnoticed
as are the events which last much longer than the whole survey.
The magnitude-limited character of the survey will not substantially bias
the detection of bright sources, but will allow only highly magnified faint
sources.
These effects leave their mark on the $u_{\rm min}$-distribution and are likely
to show preference for specific $u_{\rm min}$ ranges. The qualitative character
of such preference can be revealed only through a complete analysis
conducted by a given survey. Once necessary adjustments are applied, the
$u_{\rm min}$-distribution should be approximately box-like.
In an ideal case such distribution would not require any further correction.
The more realistic case is presented in \S 3.

\section{Basic formalism}

As mentioned in the previous section all values of $u_{\rm min}$ in the range 
between $0$ and $u^{*}_{\rm min}$ are theoretically equally likely, and, 
therefore, $u_{\rm min}$ should be box-like distributed:
\be
f(u_{\rm min}) = 
\left\{
\begin{array}{ccc} 
\frac{1}{u^{*}_{\rm min}} & {\rm if} & 0 \le u_{\rm min} \le u^{*}_{\rm min} \\ 
0 & {\rm if} & u_{\rm min}>u^{*}_{\rm min} \\
\end{array}
\right. 
\lab{uminprob}
\ee
Therefore the moments of the distribution should be given by
\be
m_n = \frac{\int_0^{\infty} \, x^n f(u_{\rm min}) \, dx}{\int_0^{\infty} \, f(u_{\rm min}) \, dx} = \frac{\int_0^{u^{*}_{\rm min}} \, x^n \frac{1}{u^{*}_{\rm min}} \, dx}{\int_0^{u^{*}_{\rm min}} \, \frac{1}{u^{*}_{\rm min}} \, dx} = \frac{{\left( u^{*}_{\rm min} \right)}^n}{n+1} \lab{moments}
\ee
As a result:
\be
E \left( \mu \left( u_{\rm min} \right) \right) \equiv m_1 = \frac{1}{2} \, u^{*}_{\rm min} \lab{uminmean}
\ee
and
\be
E \left( \sigma^2 \left( u_{\rm min} \right) \right) \equiv m_2 - m_1^2 = \frac{1}{12} \, {\left( u^{*}_{min} \right)}^2, \lab{uminvar}
\ee
where $E$ stands for the expectation value.
Now the question is how one can use this information to make a better
estimate of the microlensing parameters.
Here we are going to call a thermodynamic analogy, which we think corresponds
closely to the currently considered case.
Imagine a huge reservoir of particles which are characterized by a certain
temperature and therefore have a certain distribution of kinetic energies.
To measure some properties of these particles one could investigate
average properties of any sub-volume of this huge reservoir because such
a sub-volume is representative of the entire volume.
In practice, one would rather attach to the reservoir a small chamber
separated from the main volume by a partition.
If the partition is semi-permeable, the number of the
particles in the small chamber will increase till the experiment is finished
and the particle properties are measured.
As long as the number of particles in the small chamber is much smaller
than the number of particles in the huge reservoir, 
the chamber will be filled with particles that originate from the 
reservoir but are on average more energetic. This is because the faster 
particles are more likely to hit the partition than one would infer from their
number density. However, if one knows the original velocity distribution, one
can make an adjustment for this effect.
After this correction, the small chamber collection should be equivalent
to a randomly chosen sub-volume mentioned above and, therefore, representative
of the huge reservoir.
The most likely state of such sub-volume is the one which maximizes the 
entropy.
The state, which maximizes the entropy is the one where the distributions
of energies of particles in the reservoir and the corrected
energies of the particles in the chamber are identical.

In the case of microlensing the reservoir is filled with all the possible
events. The small chamber corresponds to a microlensing survey, which
gathers the data. The factors biasing the distribution of event parameters
are both detection efficiencies of the survey as well as blending of the 
events.
Once corrected for the known biases, the distribution of $u_{\rm min}$ of 
events in the survey should closely resemble the distribution
in the reservoir. 
And it does unless there are hidden biases associated with data reduction
and analysis.
We will define the state of maximum microlensing entropy as the one in which
the $u_{\rm min}$-distribution resembles the expected distribution to the
highest possible extent.
Because the reservoir contains a tremendous number
of possible microlensing events, its distribution of $u_{\rm min}$ is to a 
very high accuracy box-like.
We want to modify the $u_{\rm min}$-distribution of the survey to make it look 
like the expected box-like distribution. 

For simplicity, we will here present only the
conditions imposed on the mean and the variance of the distribution.
We define:
\be
u^{\rm new}_{{\rm min}, i} = g_i(P) \, u_{{\rm min}, i}, \lab{uminnew}
\ee
where $g_i(P)$ is a function of different parameters designated by the capital
$P$ (in what follows we will use just $g_i$ to avoid clutter).
The two most extreme cases are that all $g_i(P)$ are equal to 1 and so
the entire transformation is in effect an identity and that all $g_i(P)$ are
different and unrelated to each other.
For the treatment with only first two moments taken into account (flawed
in the general case as described in the Introduction),
the likelihood of certain configuration of $u_{\rm min}$ values (if no other
information is available) is
\be
L(u_{{\rm min}, i}, i \, \in \overline{1,N}) = \exp(\frac{-{\chi}^2_{\mu}}{2}) \cdot
\exp(\frac{-{\chi}^2_{{\sigma}^2}}{2}) \lab{prob}
\ee
The maximization of this probability is equivalent to minimization of
\be
-2\ln{L} = \chi^2_{\mu} + \chi^2_{{\sigma}^2}, \lab{logprob}
\ee
where
\be
\chi^2_{\mu} = \frac{ {\left( \frac{1}{N} 
\sum_{i=1}^{N} g_i \, u_{{\rm min}, i} - \frac{1}{2}
u^{*}_{\rm min} \right)}^2 }{\frac{1}{N} \left[ \frac{1}{N-1} \sum_{i=1}^{N} {\left( g_i u_{{\rm min}, i} - \frac{1}{N} \sum_{i=1}^{N} g_i u_{{\rm min}, i} \right)}^2 \right]}
\lab{chi2mean}
\ee
and
\be
\chi^2_{{\sigma}^2} = \frac{ {\left( \left[ \frac{1}{N-1} \sum_{i=1}^{N} {\left( g_i u_{{\rm min}, i} - \frac{1}{N} \sum_{i=1}^{N} g_i u_{{\rm min}, i} \right)}^2 \right] - \frac{1}{12} {\left( u^{*}_{\rm min} \right)}^2 \right)}^2 }{ \frac{ K(u_{\rm min}, g) - 1 }{N} {\left[ \frac{1}{N-1} \sum_{i=1}^{N} {\left( g_i u_{{\rm min}, i} - \frac{1}{N} \sum_{i=1}^{N} g_i u_{{\rm min}, i} \right)}^2 \right]}^2
}
\lab{chi2var}
\ee
In the general case, the kurtosis $K(u_{\rm min},g)$ is a function of both 
the original
distribution and the modifying functions and so it enters the minimization
in an active way making all the expressions very complicated.
Here we are going to consider a special case where
\be
\bigwedge_{i \in \overline{1,N}} g_i = \alpha = {\rm const}.
\lab{alpha}
\ee
Then the equations\rr{chi2mean} and\rr{chi2var} are simplified to
\be
\chi^2_{\mu} = \frac{ {\left( \alpha \mu - \frac{1}{2}
u^{*}_{\rm min} \right)}^2 }{{\alpha}^2 \, {\sigma}^2}
\lab{chi2meanalpha}
\ee
and
\be
\chi^2_{{\sigma}^2} = \frac{ {\left( {\alpha}^2 {\sigma}^2 - \frac{1}{12} {\left( u^{*}_{\rm min} \right)}^2 \right)}^2 }{ \frac{ K(u_{\rm min}) - 1 }{N}
{\alpha}^4 {\sigma}^4 },
\lab{chi2varalpha}
\ee
where $\mu$ and ${\sigma}^2$ are the mean and dispersion of the {\em original}
distribution of $u_{\rm min}$, respectively.

Now the kurtosis $K$ is only a function of the initial distribution of 
$u_{\rm min}$,
because the higher standardized moments are invariable against multiplication
of the entire distribution by a constant. Therefore, in minimization, $K$
can be treated as constant of the sought value of $\alpha$.
More importantly, all the standardized moments higher than variance remain
constant under the multiplication of the whole distribution by a constant.
In this case, the infinite number of conditions on all the moments of
the distribution is equivalent to two conditions on the mean and variance
of the distribution.
We start with a separate minimizations of equations\rr{chi2meanalpha}
and\rr{chi2varalpha} to earn some intuitive understanding of the desired
corrections to the observed set of $u_{\rm min}$ values.
From\rr{chi2meanalpha}:
\be
\frac{\partial \chi^2_{\mu}}{\partial \alpha} = \frac{ N \, u^{*}_{\rm min} \left( \alpha \mu - \frac{1}{2}u^{*}_{\rm min} \right) }{{\alpha}^3 \, {\sigma}^2}
\lab{chi2meanalphapartial}
\ee
Therefore
\be
\frac{\partial \chi^2_{\mu}}{\partial \alpha} = 0 \;\;\;\;\; \Longleftrightarrow \;\;\;\;\;     \alpha = \frac{\frac{1}{2} u^{*}_{\rm min}}{\mu}.
\lab{alphasol1}
\ee
From\rr{chi2varalpha}:
\be
\frac{\partial \chi^2_{{\sigma}^2}}{\partial \alpha} = 
\frac{ \frac{N}{3(K-1)} {\left( u^{*}_{\rm min} \right)}^2 \left( {\alpha}^2 {\sigma}^2 - \frac{1}{12} {\left( u^{*}_{\rm min} \right)}^2 \right) }{ {\alpha}^5 {\sigma}^4 }
\lab{chi2varalphapartial}
\ee
Therefore
\be
\frac{\partial \chi^2_{{\sigma}^2}}{\partial \alpha} = 0 \;\;\;\;\; \Longleftrightarrow \;\;\;\;\;     {\alpha}^2 = \frac{ \frac{1}{12} {\left( u^{*}_{\rm min} \right)}^2 }{ {\sigma}^2 }
\lab{alphasol2}
\ee
Solutions\rr{alphasol1} and\rr{alphasol2} are what one would expect. 
They force either the mean or variance of the modified distribution
to be equal to the theoretically expected values reported in\rr{uminmean}
and\rr{uminvar}.
Note that the requirement that the $\alpha$ returned by\rr{alphasol1} 
and\rr{alphasol2} be identical to each other is equivalent to ${\mu}^2 = 3 
{\sigma}^2$
(every box-like distribution with a support in the range from 0 to a constant
meets this condition).
Now we combine\rr{chi2meanalphapartial} and\rr{chi2varalphapartial}
to find the value of $\alpha$ that is optimum from the point of view of both
mean and variance of the distribution.
\be
\frac{\partial \chi^2_{\mu}}{\partial \alpha} + \frac{\partial \chi^2_{{\sigma}^2}}{\partial \alpha} = 0 \;\;\;\;\; \Longleftrightarrow \;\;\;\;\;
{\alpha}^3 + \frac{2-3(K-1)}{3(K-1)} \frac{u^{*}_{\rm min}}{\mu} {\alpha}^2 -
\frac{1}{36(K-1)} \frac{{\left( u^{*}_{\rm min} \right)}^3}{\mu {\sigma}^2} = 0
\lab{alphaequation}
\ee
We define:
\be
B \equiv \frac{2-3(K-1)}{3(K-1)} \frac{u^{*}_{\rm min}}{\mu} \;\;\;\;\;
{\rm and} \;\;\;\;\; D \equiv - \frac{1}{36(K-1)} \frac{{\left( u^{*}_{\rm min} \right)}^3}{\mu {\sigma}^2}
\lab{bddef}
\ee
We introduce a new variable $y = \alpha + \frac{B}{3}$. Then the 
equation\rr{alphaequation} takes the form:
\be
y^3 - \frac{B^2}{3} \, y + \left( \frac{2B^3}{27} + D \right) = 0
\lab{yequation}
\ee
From the theory of solving third order equations one knows that
equation\rr{yequation} has two complex and only one real solution if
\be
\Delta \equiv D \left( \frac{B^3}{27} + \frac{D}{4} \right) > 0
\lab{onesolcond}
\ee
If $K>1$ then we see from\rr{bddef} that $D<0$. Therefore, 
condition\rr{onesolcond} is equivalent to 
\be
\left( \frac{B^3}{27} + \frac{D}{4} \right) < 0
\lab{Kmore1onesolcond}
\ee
Condition\rr{Kmore1onesolcond} is true, independent of the values of $\mu$
and ${\sigma}^2$, if
\be
B \leq 0 \;\;\;\;\; \Longleftrightarrow \;\;\;\;\; K \geq \frac{5}{3}
\lab{kurtosiscond}
\ee
Condition\rr{kurtosiscond} is likely to be the case most of the 
time\footnote{The kurtosis of the box-like distribution is $K=1.8$, so most 
of the deviations from this distribution will have $K \geq \frac{5}{3}$.}  
and the only real root of equation\rr{alphaequation} will be the most probable 
solution for $\alpha$. This solution can be written as
\be
\alpha = \sqrt[3\,]{-\frac{B^3}{27}-\frac{D}{2} + \sqrt{\Delta}} \; + 
\sqrt[3\,]{-\frac{B^3}{27}-\frac{D}{2} - \sqrt{\Delta}} \; - \frac{B}{3}
\lab{alphasol}
\ee
In the case when all the roots of equation\rr{alphaequation} are real
one has to choose the one which minimizes the combined ${\chi}^2$ of
equation\rr{logprob} instead of just taking the solution\rr{alphasol}.
We do not discuss such complications in any detail because the procedure
is mathematically straightforward and completely standard.
 
Multiplication of the entire $u_{\rm min}$-distribution by $\alpha$ is likely
to change the number of events in the range $(0, u_{\rm min}^{*})$.
The described procedure should be repeated on new sets as many times
as necessary for the results to converge.
Therefore, the result of the analysis can be given as
$\alpha = \prod_{i=1}^{R} \alpha_{i}$, where $R$ is the number of needed
repetitions and $\alpha_{i}$ are corrections of individual iterations.

In Figure 2 we present the application of equation\rr{alphasol} to an 
artificial
set of 17 $u_{\rm min}$ values. The dashed line is a theoretically
expected distribution for magnification threshold of $A_{\rm max}^{*} = 1.34$
or $u^{*}_{\rm min} = 1.0$. The thin solid line is an original, uncorrected
cumulative distribution of $u_{\rm min}$ values. The thick solid line
is this distribution multiplied by $\alpha = 0.9$, as obtained from 
equation\rr{alphasol}. The improvement is easily visible and can be quantified
by reduction in maximum vertical distance between the two distribution $D_N$
(formally introduced in the next section).

\section{Remarks about the more general treatment and impossibility to 
constrain individual blended events}

We will now consider the most general
case as described by equations\rr{chi2mean},\rr{chi2var}, and the infinite
number of other equations for the higher moments.
Because all the moments now contain the statistical information, the only 
solution is to turn to the original Kolmogorov treatment.

The Kolmogorov test verifies the hypothesis that a continuous variable X 
could have been drawn from a cumulative distribution function $F_0(x)$. 
The test statistic is
\be
D_N = \sup_x \left| F_0(x)-S_N(x) \right|,
\lab{DN}
\ee
where $S_N(x)$ is the observed cumulative distribution function based on
the ordered sample of N measurements. If $d_N(1-\beta)$ designate 
quantiles of the statistic $D_N$, then $P(D_N \geq d_N(1-\beta))= \beta$.
The higher $\beta$, the less sensitive the test is.

Suppose that we construct all the possible sets of corrected 
$u_{\rm min}$ values from the interval 0 to $u^{*}_{\rm min}$.
Then for each set we can find a $D_N$ statistic and a corresponding 
significance level $\beta$.
In a formal Kolmogorov-type analysis, it is inappropriate to conclude
that the relative probabilities of different sets scale as
$
\frac{P({\rm set 1})}{P({\rm set 2})}= \frac{\beta({\rm set 2})}{\beta({\rm set 1})}
$. 
However, even though we cannot assign relative probabilities of
certain corrected configurations, we can select the set of 
$u_{\rm min}$ values with smallest $\beta$ as the preferred solution
(for the cumulative distribution only).

The main problem with this approach is that from the point of view of 
the cumulative distribution only, there will be a number of optimum solutions,
all of which, except one, will mix the original order of $u_{\rm min}$ values.
To avoid total chaos one can consider a simplified situation when: 
\renewcommand{\theenumi}{\Alph{enumi}}
\begin{enumerate}
\item the order of $u_{\rm min}$ to be corrected is going to be conserved,
that is that the event with the i-th lowest $u_{\rm min}$, will remain
the event with the i-th lowest $u_{\rm min}$ after the correction has
been done.
\end{enumerate}
Imagine that we operate in this idealized situation and select $N$ microlensing
events and order them according to increasing $u_{\rm min}$.
One will reach the highest possible agreement if one sets:
\be
u_{{\rm min},i} = \frac{i-0.5}{N+1} \; u^{*}_{\rm min}. \lab{theoumin}
\ee
If correcting factors $g_i$ can be represented as a function of 
$u_{{\rm min}, i}$ that is $g_i = g(u_{{\rm min}, i}$), then as long as 
$\partial g/\partial u_{\rm min} > -1/u_{\rm min}^2 $ in the range 
$(0,u_{\rm min}^{*})$, condition A. does not have to be imposed but is 
naturally met.
Condition A. makes the most sense if one deals with a separated set
of events. ``Separated set'' here is a well defined concept described
as
\be
\bigwedge_{i \in \overline{1,N}} \delta u_{{\rm min},i} \ll {\rm min} \left([u_{{\rm min},(i+1)}-u_{{\rm min},i}], \;[u_{{\rm min},i}-u_{{\rm min},(i-1)}]\right),
\lab{separatedset}
\ee
where $\delta u_{{\rm min},i}$ is the formal error in $u_{{\rm min},i}$.
At the first glance, one may think that due to the requirement of small
formal errors, the inequality\rr{separatedset}
automatically guarantees that all $u_{\rm min,i}$ were determined correctly.
However, that may not be the case, when unrecognized biases are present.
In a separated set case, the constraints from the fits to individual events
are crucial in establishing the order of $u_{\rm min}$ values and the 
Kolmogorov test is used to stretch or squeeze the entire distribution.
The prescription described here and in \S 3 can be implemented 
in an iterative program, which can converge on the solution using both
the cumulative distribution information as well as other constraints 
on individual $u_{{\rm min}, i}$ with the relative weights selected
by the investigator.

It is important to note that one is not allowed to draw the same 
statistical conclusions by fixing the $u_{\rm min}$-positions of some events 
and ``filling the gaps'' with the others. 
To give an example, one may hope that the information contained in the 
microlensing character
of the ensemble can help to deblend the events with highly degenerate
fits. Imagine that we have $(N-1)$ events with nicely determined parameters
(for which we would be happy to put $g_i \equiv 1$) and one blending event
with a very uncertain blending fraction (let say k-th on the ordered list).
This corresponds to a situation when only one $g_i \neq 1 = \alpha$.
In this case, the blindly applied Kolmogorov analysis is going to return 
quite a strong 
preference for a certain value of $u_{\rm min, k}$ of this particular event. 
However, the actual probability that the considered event will be in this 
narrow preferred range is low and all the values of $u_{\rm min}$ between 0 and 
$u^{*}_{\rm min}$ are almost equally likely. This is because the right treatment 
here is just to consider a conditional probability of a certain $u_{\rm min}$ 
value given the set of $(N-1)$ other $u_{\rm min}$ values. Because the reservoir 
of possible microlensing events is huge, this conditional probability is 
extremely close to the flat distribution expected from the infinite set.
Therefore, the rule is to either adjust the parameters of all events 
applying the Kolmogorov approach (and extra constraints) to the whole set of 
$u_{\rm min}$ values or leave it untouched.
All the naive ``partial'' procedures involving Kolmogorov test
will produce spurious results.

\section{Summary}

We have investigated the question how the information that events constitute
microlensing family can help to reduce systematic errors in parameter
determination.
We have argued that the Kolmogorov test for the agreement between the
observed and theoretical cumulative distributions of the impact parameter
$u_{\rm min}$ can put useful additional constraints on microlensing parameters.
However, there is no natural way to incorporate the classical Kolmogorov
approach into the convenient maximum likelihood analysis.
Fortunately, the constraints on the cumulative distribution can be replaced 
with the constraints on the infinite number of the moments of the distribution.
We explain why taking into account only a few lower moments of the 
distribution is flawed, and, therefore eliminates the usefulness of such
limited moment approach. 
However, we show that when the whole distribution of $u_{\rm min}$ values
is multiplied by the same factor $\alpha$, all the information about the 
probabilities of different $\alpha$'s is stored in the mean and variance
of the original distribution and the selection criteria of events.
Assuming that no other statistical information on events is available
(or relevant at this point), we solve for $\alpha$. Our results are
given in equation\rr{alphasol} of \S 3.

In the follow-up, given the corrected $u_{\rm min}$ values one may
keep them fixed and re-determine the other parameters. Such approach is
applicable if there is a systematic homogeneous bias of all $u_{\rm min}$
values.
The fully consistent treatment
should proceed through the simultaneous determination of the multiplying
factor $\alpha$ and the parameters of individual events.
A useful consistency check would be to select different $u_{\rm min}^{*}$
values and make sure that $\alpha$ does not strongly depend on this choice
\footnote{Here we mean $u_{\rm min}^{*}$ values that do not differ too much
from the original one --- $\alpha$ is a local correction and as such is
expected to very if $u_{\rm min}^{*}$ changes substantially.}.

One may ask when it is justified to apply an $\alpha$-type correction
to the data. This is a serious concern because it is normal that every actual 
realization of $u_{\rm min}$-distribution will be affected by finite number 
statistics and, as a result, $\alpha$ is expected to be $\neq 1$ even for 
unbiased data, which need no correction. However, in this case, correction by 
$\alpha$ will typically be small and will not significantly
change the properties of the distribution. Consequently, it seems safe
to apply the $\alpha$-type correction even if the original distribution
is consistent with the expected one.

Finally, we should examine the assumption that the microlensing sample
under analysis is clean.
In the traditional approach of massive
microlensing searches directed at random events, it is very hard to 
construct a completely clean microlensing sample. 
In defense of $\alpha$-type correction, we should notice, that sensitivity to 
contamination is common to 
all determinations associated with microlensing. For example:
1) few long-duration variables mistakenly classified as microlenses
   can change the microlensing optical depth by a factor of two,
2) few very short stellar flares mistakenly classified as
   microlenses can entirely change the slope and cutoff of the mass
   function of lenses derived from the duration distribution of
   events.\\
Similarly our method may bias the final $u_{\rm min}$-distribution if some
candidate events are not microlenses. 
Therefore, a particular survey should apply this method to their data
only if the bias caused by sample contamination is smaller
than the advantage gained due to the described adjustment.
A potential improvement can be judged using a learning set of 
highly-constrained events\footnote{Such a learning set can also answer the 
question of what type of systematic errors are best removed with
$\alpha$-type correction.
However, in general, the understanding of the nature of systematic
errors is not required to achieve improvement in parameter determination.}.

However, $u_{\rm min}$-test is often used as evidence for microlensing 
character of the events. This makes its status somewhat different from
the optical depth or duration distribution of the events.
As a result, what we need in an ideal case is a way to separate the 
``convoluted'' nature of $u_{\rm min}$-test (which intrinsically allows to 
check both the clean character of the sample and the correctness of 
$u_{\rm min}$ values) into two independent issues. 
If we could select a clean sample, then we could limit the role of 
$u_{\rm min}$-test to a single task of parameter control. In this way
$u_{\rm min}$-distribution could become an active, unbiased tool in 
parameter determination, as we have presented it above.

Here we give two examples of realistic situation, when the clean
character of the sample can be established with very high
probability.\\[0.1cm]
{\em 1. Events on demand from high proper motion lenses.}\\
In the current microlensing searches there is no expectation of 
the location of the next event. The next event happens somewhere on
one of $\sim 10^7$ monitored stars. In this scenario and with typical
optical depths of order of $10^{-7}-10^{-6}$, it is expected
that most varying objects will not be microlensing events. Moreover,
there will be enough variables detected per each microlensing event to produce
cases that look very much like microlensing and can contaminate the sample. 
Fortunately, observations do not have to follow this pattern.
Some high proper motion stars toward dense stellar fields 
(e.g., Galactic bulge) will within $\sim 10$ years become sufficiently 
aligned with the known bulge sources to produce microlensing events 
(Salim \& Gould 2000; Drake et al.\ 2002). 
As the occurrence of such events can be predicted based on known stellar
motions, one can check the sky patch under investigation for possible 
non-microlensing contaminants in advance. 
More importantly, if the event happens when scheduled, the probability that
it is not microlensing is negligible.\\[0.1cm]
{\em 2. Astrometrically constrained events.}\\
Single variable stars (including cataclysmic ones) are the main 
contaminants of microlensing samples. With high accuracy astrometric
observations of candidate events carried out by Space Interferometry
Mission (SIM) or similar astrometric satellite, it 
is easy to eliminate such contaminants: the centroids of varying flux
due to variables do not move, whereas the centroids of light from 
microlensing images follow a well defined path (e.g., Boden et al.\ 1998). 
A detection of the centroid shift of varying flux (even the one 
insufficient to constrain microlensing parameters from astrometry
alone) will be a very good indication of the microlensing character of
the event.\\[0.3cm]
Both types of samples described above should be clean enough for the 
purpose of non-restrictive application of $\alpha$-type correction,
which may be based on light curves alone.

\acknowledgments

We thank Thor Vandehei whose genuine desire to accurately de-bias
the MACHO data on the Large Magellanic Cloud triggered this investigation.
We are grateful to Kim Griest for stimulating discussions on the issues
related to this project.
This work was performed under the auspices of the U.S. Department of Energy 
by University of California Lawrence Livermore National Laboratory under 
contract No. W-7405-Eng-48.

\clearpage

\begin{figure}[htb]
\includegraphics[width=17cm]{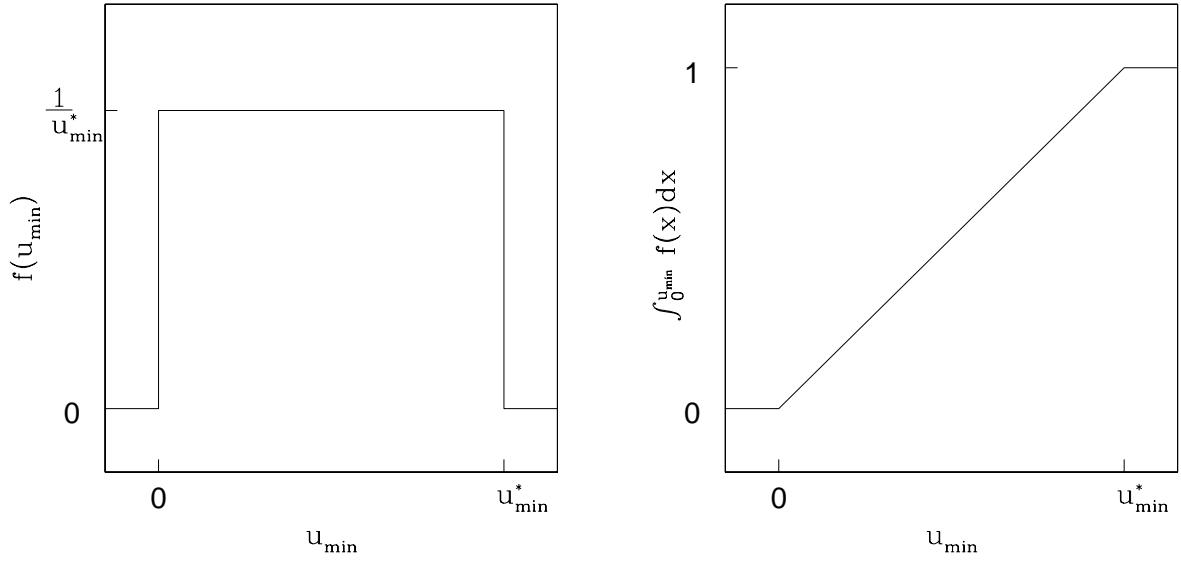}
\caption{
The expected distribution of impact parameter
$u_{\rm min}$ for microlensing with a minimum magnification at maximum light
bigger than 1. The left panel shows the probability that an event
will have a certain $u_{\rm min}$ value. Drop to 0 at $u_{\rm min}^{*}$ corresponds
to amplification threshold introduced by survey's selection criteria.
The right panel presents the corresponding cumulative distribution function.
This is the shape which is referred to as the theoretically expected cumulative
distribution for the microlensing Kolmogorov test.
}
\end{figure}

\clearpage

\begin{figure}[htb]
\includegraphics[width=15cm]{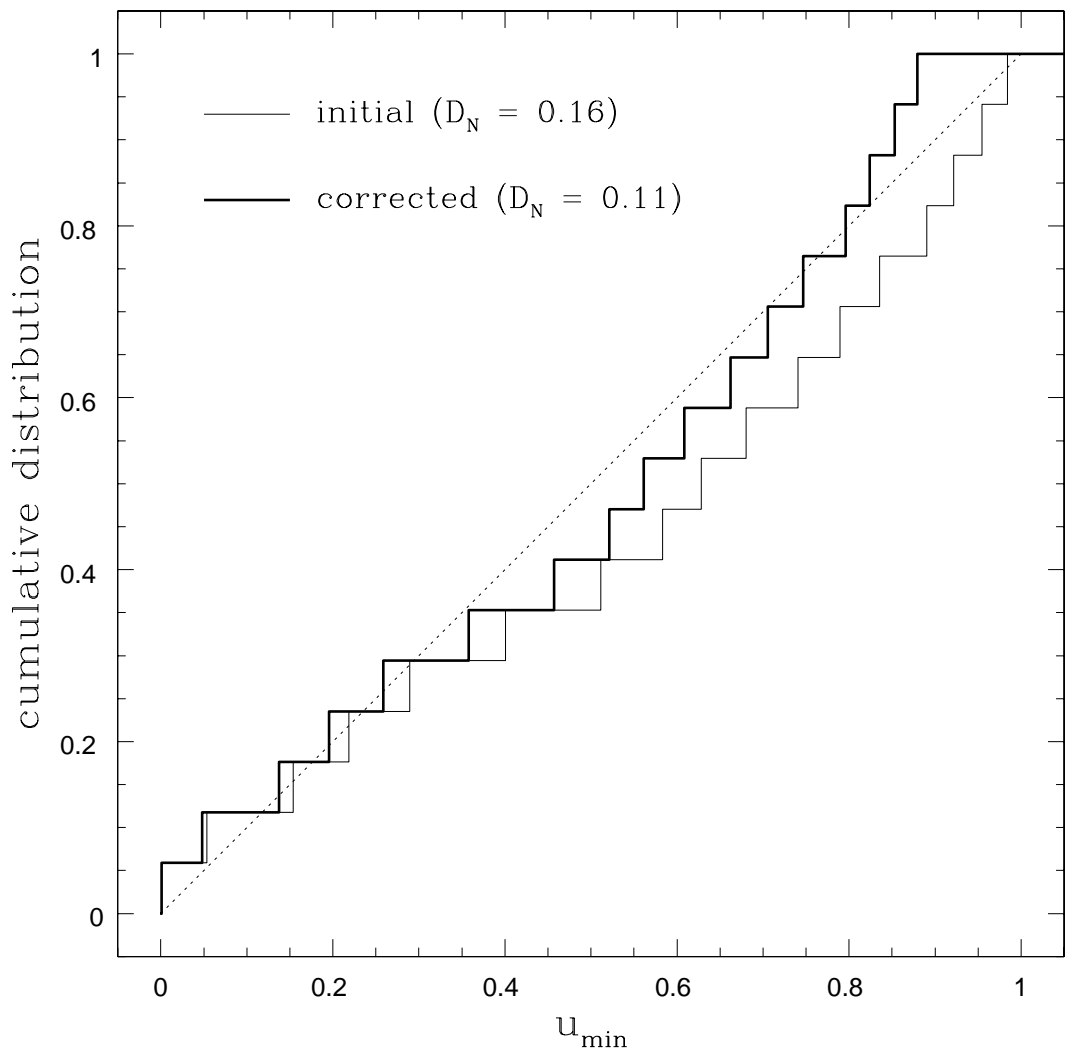}
\caption{
Correction of the original $u_{\rm min}$
cumulative distribution as applied to an artificial
set of 17 $u_{\rm min}$ values. The {\em dashed line} is a theoretically
expected distribution for magnification threshold of $A_{\rm max}^{*} = 1.34$
or $u^{*}_{\rm min} = 1.0$ in the case of infinite number of events. 
The {\em thin solid line} is the original, uncorrected
cumulative distribution of $u_{\rm min}$ values. The {\em thick solid line}
is the corrected distribution, i.e. the original distribution multiplied 
by $\alpha = 0.9$. The correction results in substantial improvement
in Kolmogorov statistic $D_N$.
}
\end{figure}

\clearpage

\end{document}